\documentclass[12pt]{article}
\usepackage{graphicx}
\usepackage{amssymb}
\usepackage{natbib}
\newcommand{\be}{\begin{equation}} \newcommand{\ee}{\end{equation}}
\newcommand{\ba}{\begin{eqnarray}} \newcommand{\ea}{\end{eqnarray}}
 \renewcommand{\bf}{\textbf}

\begin{document}

\title{CMB  Polarization  and Temperature Power Spectra Estimation
using Linear Combination of WMAP 5-year Maps}
\author{ Pramoda Kumar Samal$^{1}$, Rajib Saha$^{2}$, 
Jacques Delabrouille$^{3}$,\\
Simon Prunet$^{4}$, Pankaj~Jain$^{1}$,
   Tarun Souradeep$^5$}
\maketitle

\begin{center}
{$^{1}$Department of Physics, Indian Institute of Technology, Kanpur, India \\ }
{$^{2}$Jet Propulsion Laboratory, M/S 169-327, 4800 Oak   Grove Drive, Pasadena,CA 91109; California Institute of Technology, Pasadena,\\ CA 91125, USA\\ }
{$^{3}$CNRS, Laboratoire APC, 10, rue Alice Domon et L\'{e}onie Duquet, 75205 Paris,
France\\}
{$^{4}$Institut d'Astrophysique de Paris, 98 bis Boulevard Arago, F-75014 Paris, France\\}
{$^{5}$Inter-University Centre for
  Astronomy and Astrophysics, Post Bag 4, Ganeshkhind, Pune 411007,
    India\\}
\end{center}

\begin{abstract}
We estimate CMB polarization and temperature power spectra using WMAP
$5$-year foreground contaminated maps. The power spectrum is estimated
by using a model independent method, which does not utilize directly the
diffuse foreground templates nor the detector noise model. The method
essentially consists of two steps, (i) removal of diffuse foregrounds
contamination by making linear combination of individual maps in
harmonic space and (ii) cross-correlation of foreground cleaned maps
to minimize detector noise bias. For temperature power spectrum we
also estimate and subtract residual unresolved point source
contamination in the cross-power spectrum using the point source model
provided by the WMAP science team.  Our $TT$, $TE$ and $EE$ power
spectra are in good agreement with the published results of the WMAP
science team. The error bars on the polarization power spectra,
however, turn out to be smaller in comparison to what is obtained by the
WMAP science team. We perform detailed numerical simulations to test for bias in
our procedure. We find that the bias is small in all
cases. A negative bias at low $l$ in $TT$ power spectrum has been
pointed in an earlier publication.  We find that the bias corrected
quadrupole power ($l(l+1)C_l/2\pi)$ is $532$ $\mu K^2$, approximately
$2.5$ times the estimate ($213.4$ $\mu K^2$) made by the WMAP team.
\end{abstract}

\section{Introduction}
The anisotropies of the Cosmic Microwave Background (CMB) radiation are 
the most important evidence behind the tiny fluctuations that are
generated by the inflationary paradigm of the Big-Bang
cosmology~(\cite{Starro82,Guth82,Bardeen83}).  One can determine
cosmological parameters precisely by measuring these
anisotropies~(\cite{Jung96a,Jung96b,Bond97,ZalSpSel97}). These
anisotropies possess a certain degree of linear polarization due to
the quadrupolar temperature pattern seen by the moving electrons in
the primordial plasma~(\cite{Rees68,Basko80}).  Recently, the WMAP
satellite has mapped the total intensity and polarization of the CMB
anisotropies over the full sky in its $5$ frequency bands from $23$
GHz to $94$ GHZ with unprecedented resolution and
sensitivity~(\cite{Bennett03a,Bennett03b,Page07,Kogut07,Hinshaw08}).
The polarization power spectrum acts as a complement to the
temperature power spectrum. It leads to better constraints on the
cosmological parameters and is also useful to break degeneracies
among certain cosmological parameters, e.g. epoch of reionization
and scalar to tensor ratio~(\cite{Kinney98}). Furthermore it has been
argued that CMB polarization may serve as a direct probe of
inflation~(\cite{Spergel97}), can test if the parity symmetry is
preserved on the cosmological scales~(\cite{Lue99,Komatsu08}), can
provide information about the epoch when the first stars begin to
form~(\cite{Critteden93,Nga95}) and provide a measure of the gravity
waves that are generated by
inflation~(\cite{Harari93,Crittenden95,Kako98}). The WMAP team have
produced their temperature and polarization power spectrum based upon
the foreground cleaned maps which are obtained using prior models of
the synchrotron, dust and free-free
components~(\cite{Kogut07,Page07}). Though this method allows one to
use all the available information about the foreground components it
is also a very important scientific task to perform an independent
analysis of the data by techniques which do not rely upon explicit
foreground modeling.

A multipole based approach for foreground removal was first proposed
by~\cite{TeEfs1996} and was implemented on the WMAP data
by~\cite{Tegmark2003}. Later, \cite{Saha2006,Saha2008,Souradeep2006}
extended this method to extract the temperature anisotropy power
spectrum of the Cosmic Microwave Background (CMB) radiation from the
raw WMAP data. The power spectrum is obtained by forming several
cleaned maps using subsets of the available maps and thereafter
cross-correlating the resulting maps.  This Internal Power Spectrum
Estimation (IPSE) method utilizes CMB data as the only input without
making any explicit modeling of the diffuse galactic foreground
components or detector noise bias.  The foreground components are
removed using the fact that in thermodynamic temperature unit, the CMB
signal is predicted to be independent of frequency since it follows a
black-body spectrum~(\cite{Mather94,Fixsen96}), while the foreground
components are frequency dependent. The detector noise bias is removed
by cross-correlating different foreground cleaned maps obtained by
using independent detector subsets. This substantially removes the
noise bias since, to a good approximation, WMAP detector noise is
uncorrelated for two different
detectors~(\cite{Jarosik03,Jarosik07}). The final power
spectra~(\cite{Saha2006,Saha2008}) obtained by IPSE agrees well with
the results published by the WMAP science team. Thus the method serves
as an independent technique to verify the main power spectrum result
obtained by the WMAP science team starting from the stage of diffuse
foreground components removal. The method has several 
advantages. First, the foreground components removal method is
entirely independent of the foreground template models. Therefore the
foreground cleaned maps are not susceptible to systematic errors that
might arise in template based methods due to incorrect template
modeling. Second, the cleaned power spectrum can be studied
analytically in the special case of full sky one iteration foreground
cleaning~(\cite{Saha2008}). This allows us to quantify and understand
the statistical properties of the residuals in the cleaned power
spectrum.  This may be very useful in the case of noisy data or when
the total number of available frequency bands are less than the total
number of independent parameters required for satisfactory modeling of
all dominant underlying components.  A detailed analytical study of
the bias in the cleaned power spectrum is presented
in~\cite{Saha2008}. Third, it is possible to obtain a model
independent estimate of the map and power spectrum of the total
foreground emission at each of the frequency bands~(\cite{ghosh2009}).

In the present paper we extract CMB polarization $EE$, $TE$ as well as
the $TT$ power spectra using the WMAP $5$-year foreground and detector
noise contaminated maps as input to IPSE -- our combined foreground
removal and power spectrum estimation procedure. Since the CMB
polarization signal is weak the polarized maps published by the WMAP
science team are dominated by the foreground components and detector
noise. This seriously limits the accuracy with which the polarized CMB
power spectrum can be extracted. However, since our method does not
use any template model to remove foreground components we argue that
our power spectrum is free from systematic effects that might arise
due to incorrect modeling of polarized (and temperature) foreground
templates.

The error bars as well as the bias in the extracted polarization power
spectrum are estimated by numerical Monte Carlo simulations. Here we
make use of explicit foreground and detector noise models. In the case
of temperature power spectrum a similar analysis reveals the presence
of a negative bias at low multipoles~(\cite{Saha2008}).  The bias
corrected temperature spectra explains almost all of the low power
observed in case of quadrupole. In this case it is also possible to
analytically obtain an estimate of the bias in some special cases.

Alternate approaches of CMB power spectrum estimation have been
studied by several authors, e.g., using foreground cleaned maps
provided by the WMAP science
team~(\cite{Fosalba04,Patanchon05,Eriksen07,Eriksen07a}), as well as
using uncleaned maps where some models of foregrounds and (or)
detector noise are necessary~(\cite{Eriksen08_a,Eriksen08_b}). Other
approaches for foreground cleaning, using needlet
coefficients~(\cite{Delabrouille08}) and harmonic variance
minimization~(\cite{Kim08_a,Kim08_b}) have also been proposed.

The organization of our paper is as follows. We describe the methodology 
of power spectrum estimation in Section~\ref{Method}. We present the 
power spectrum results along with their covariance structure in Section~\ref{Result}.
Finally, we conclude in Section~\ref{Conclusion}.

\section{Method}
\label{Method}
The basic procedure for extracting the temperature power spectrum is described
in~\cite{Saha2006} and~\cite{Saha2008}.  Here we generalize this to include 
polarization. The basic maps for the case of polarization are available in terms 
of the Stokes parameters $Q$ and $U$. Since these are coordinate dependent quantities
it is more convenient to work with the coordinate independent $E$ and $B$ modes~(\cite{Zaldarriaga97,Seljak97,Kaminkowski97}). 
Another problem with 
using $Q$ and $U$ maps is that the $E$ and $B$ modes mix with
one another when one applies  a sky mask~(\cite{Jaffe2000,Tegmark2001,Bunn2003,SmithZal2007})
to remove heavily contaminated Galactic regions. This demands an extra data processing step 
to isolate the actual CMB $E$ and $B$ mode power spectra from their mixture. To avoid this problem we start
by converting full sky $Q$ and $U$ maps to full sky $E$ and $B$ maps and apply mask
whenever required on the resultant maps. This is
similar to what is proposed in~\cite{Betoule09} for  
estimating $r=T/S$ for the Planck satellite mission and the
Experimental Probe of Inflationary Cosmology
(EPIC). To obtain the $E$ and $B$ maps  
we first expand the full sky spin $\pm 2$ fields $(Q\pm iU)$ in terms of spin-2 
spherical harmonics $ \;_{\pm2}Y_{lm}(\hat n)$
\begin{eqnarray}
(Q+iU)(\hat n)=\sum_{lm} a_{2,lm}\;_2Y_{lm}(\hat n) \nonumber \\
(Q-iU)(\hat n)=\sum_{lm} a_{-2,lm}\;_{-2}Y_{lm}(\hat n)
\end{eqnarray}
Since both  $Q$ and $U$ are real, one can show that the expansion coefficients 
obey $a^*_{-2,lm}=a_{2,l-m}$. The spin-0 $E$ and $B$ are now obtained by the 
usual spherical harmonic transform,
\begin{eqnarray}
E(\hat n)=\sum_{l\ge 2,|m|\le l} a_{lm}^E Y_{lm}(\hat n) \nonumber \\
B(\hat n)=\sum_{l\ge 2,|m|\le l} a_{lm}^B Y_{lm}(\hat n)
\end{eqnarray}
where
\begin{eqnarray}
a_{lm}^E=\frac{1}{2}(a_{2,lm}+a_{-2,lm}) \nonumber \\
a_{lm}^B=\frac{1}{2i}(a_{2,lm}-a_{-2,lm})
\end{eqnarray}
This gives us $10$ different full sky maps for each of the $E$ and $B$ fields corresponding
to the $10$ WMAP Differencing Assemblies (DAs). The 10 DAs are labeled as K, Ka, Q1, Q2, V1, V2,
W1, W2, W3, W4 corresponding to the five different frequency channels
K, Ka, Q, V and W. We note that the bands Q, V and W have 2, 2, 4 DAs 
respectively.

We first eliminate the highly contaminated Galactic plane from all the $10$ DA maps 
using P06 mask~(\cite{Page07}).
 This procedure is slightly different from
that  described in~\cite{Saha2006,Saha2008}. In the latter   
case the authors cleaned the entire unmasked sky in nine iterations and also
produced a full sky
foreground cleaned temperature maps. However to eliminate potential residual foreground 
contamination arising from the Galactic plane the KQ85 mask~(\cite{Gold08}) is applied before computing the power 
spectrum. In the present work, 
we apply the mask right at the beginning since we are interested in extracting
only the power spectrum. 
To cross-check our one iteration
method we also divide the EE maps in several parts and then perform foreground 
removal in the iterative approach. We find that, the final power spectrum of this method 
is similar to the one iteration case.

We select different possible linear combinations of $4$ maps out of
the available DAs as described below.  The entire set of linear
combinations are listed in Table \ref{tab:combination}.  Each of these
linear combinations independently lead to a clean map.

The cleaning is accomplished independently for each $l$, by linearly combining these maps with weights,
$\hat w_l^a$, such that the spherical harmonic components of the cleaned map are 
given by,
\begin{equation}
a_{lm}^{\rm Clean}=\sum_{a=1}^{n_c} \hat w_l^{a}\frac{a_{lm}^a}{B_l^a} \, .
\label{c_map}
\end{equation}
Here $n_c$ is the total number of maps used for cleaning. In the present
case of $4$ channel cleaning, $n_c = 4$. The factor $B_l^a$ is the
circularized beam transform function for the  frequency band  $a$~(\cite{Hill08}). The 
weights $\hat w_l^a$ are chosen so as to minimize the
total power subject to the constraint
\begin{equation}
 {\hat {\bf W}}_l{\bf e}_0={\bf e}^T_0{\hat{\bf W}}^T_l=1 \, ,
\label{constraint}
\end{equation}
where 
 ${\bf e}_0$  is a column vector with unit elements
 \begin{equation}
 {\bf e}_0 =\left(
 \begin{array}{c}
 1  \\
 ..  \\
 ..  \\
 1
 \end{array}
 \right) \, ,
 \end{equation}
and $\hat {\bf W}_l$ is the row vector $(\hat w_l^1,\hat w_l^2,..,\hat w_l^{n_c})$.
This constraint is required so as to preserve the CMB signal. The weights are 
obtained using the empirical covariance matrix,  $\hat{\bf C}_l$, by
the relationship,~(\cite{Saha2006,Saha2008,Tegmark2003,TeEfs1996,Eriksen_ilc,Del_Cardoso})
\begin{equation}
\hat {\bf W}_l =\frac{{\bf e}^T_0 (\hat{\bf C}_l)^{-1}}{{\bf e}^T_0(\hat{\bf C}_l)^{-1}{\bf e}_0}\, .
\end{equation} 

We label the resulting cleaned maps as C{\bf i} and CA{\bf i}
where ${\bf i}=1,2,...,24$. Here the maps C{\bf i} use the DAs  
K along with possible combinations of DAs from the bands Q, V and W. 
Similarly the maps CA{\bf i} include Ka instead of K. The entire 
nomenclature is listed in Table \ref{tab:combination}.
In the case of the W band we average over two DAs before we start the 
foreground cleaning. Hence in Table \ref{tab:combination} the notation 
W12, for example, refers to the average of the DAs W1 and W2. This
averaging is not essential to the procedure and one may also directly use the 
original WMAP DAs. However averaging leads to a  reduced detector noise in
each cleaned map. 
\begin{table}
\begin{tabular}{|l |l|}
\hline                        
K + Q1 + V1 + W12 = C1 &  Ka  + Q1 + V1 + W12 = CA1\\
                              \hline
K + Q1 + V1 + W13 = C2  &  Ka  + Q1 + V1 + W13 = CA2\\
                              \hline
K + Q1 + V1 + W14 = C3  &  Ka  + Q1 + V1 + W14 = CA3 \\
                              \hline
K + Q1 + V1 + W23 = C4  &  Ka  + Q1 + V1 + W23 = CA4 \\
                              \hline
K + Q1 + V1 + W24 = C5  &  Ka  + Q1 + V1 + W24 = CA5\\
                              \hline
K + Q1 + V1 + W34 = C6  &  Ka  + Q1 + V1 + W34 = CA6 \\
                              \hline
                              \hline
K + Q2 + V2 + W12 = C7  &  Ka  + Q2 + V2 + W12 = CA7 \\
                              \hline
K + Q2 + V2 + W13 = C8   & Ka  + Q2 + V2 + W13 = CA8 \\
                              \hline 
K + Q2 + V2 + W14 = C9   & Ka  + Q2 + V2 + W14 = CA9 \\ 
                              \hline 
K + Q2 + V2 + W23 = C10  &  Ka  + Q2 + V2 + W23 = CA10\\ 
                              \hline 
K + Q2 + V2 + W24 = C11  &  Ka  + Q2 + V2 + W24 = CA11\\ 
                              \hline 
K + Q2 + V2 + W34 = C12  &  Ka  + Q2 + V2 + W34 = CA12\\ 
                              \hline 
                              \hline 
K + Q1 + V2 + W12 = C13  &  Ka  + Q1 + V2 + W12 = CA13 \\ 
                              \hline 
K + Q1 + V2 + W13  = C14  &  Ka  + Q1 + V2 + W13  = CA14\\ 
                              \hline 
K + Q1 + V2 + W14  = C15  &  Ka  + Q1 + V2 + W14  = CA15\\ 
                              \hline 
K + Q1 + V2 + W23  = C16  &  Ka  + Q1 + V2 + W23  = CA16\\ 
                              \hline 
K + Q1 + V2 + W24  = C17  &  Ka  + Q1 + V2 + W24  = CA17\\ 
                              \hline 
K + Q1 + V2 + W34  = C18  &  Ka  + Q1 + V2 + W34  = CA18\\ 
                              \hline 
                              \hline 
K + Q2 + V1 + W12  = C19   &   Ka + Q2 + V1 + W12  = CA19 \\ 
                              \hline 
K + Q2 + V1 + W13  = C20  &  Ka  + Q2 + V1 + W13  = CA20\\ 
                              \hline 
K + Q2 + V1 + W14  = C21  &  Ka  + Q2 + V1 + W14  = CA21\\ 
                              \hline 
K + Q2 + V1 + W23  = C22  &  Ka  + Q2 + V1 + W23  = CA22\\ 
                              \hline 
K + Q2 + V1 + W24  = C23  &  Ka  + Q2 + V1 + W24  = CA23\\ 
                              \hline 
K + Q2 + V1 + W34 = C24   & Ka  + Q2 + V1 + W34 = CA24 \\ 
 \hline 
\end{tabular} 
 \caption{ List of the different combination of the DA maps, used to obtain the final 48 cleaned maps, denoted by $C{\bf i}$ and $CA{\bf i}$ where ${\bf i}$ = 1, 2, \ldots, 24.}
\label{tab:combination}
\end{table}                                                                  

After obtaining the $48$ cleaned maps we cross-correlate them in selected 
combinations in order to reduce the contribution due to detector noise. 
We cross-correlate
all pairs of maps such that the two cleaned maps in each pair are formed
by distinct DAs. This gives us $24$ cross-correlated power spectra on the
masked sky.
All the possible cross-correlations are listed in Table~\ref{tab3}. 
\begin{table}
\small
\centering
\begin{tabular}{|l |l|l|l|l|l|}
\hline
                   &                     &                    &                   &                     &  \\
C1 $\otimes$ CA12  &  C2 $\otimes$ CA11  &  C3 $\otimes$ CA10 & C4 $\otimes$ CA9  &  C5 $\otimes$ CA8   &  C6 $\otimes$ CA7\\
C7 $\otimes$ CA6   &  C8 $\otimes$ CA5   &  C9 $\otimes$ CA4  & C10$\otimes$ CA3  &  C11$\otimes$ CA2   &  C12$\otimes$ CA1\\
C13$\otimes$ CA24  &  C14$\otimes$ CA23  &  C15$\otimes$ CA22 & C16$\otimes$ CA21 &  C17$\otimes$ CA20  &  C18$\otimes$ CA19\\
C19$\otimes$ CA18  &  C20$\otimes$ CA17  &  C21$\otimes$ CA16 & C22$\otimes$ CA15 &  C23$\otimes$ CA14  &  C24$\otimes$ CA13\\
                   &                     &                    &                   &                     &  \\
\hline
\end{tabular}
\caption{List of all the $24$ cross-power spectra using the $48$ cleaned maps,
C1, C2,..., C24 and CA1, CA2,..., CA24.}
\label{tab3}
\end{table}

We convert each of the $24$ masked sky power spectra into full sky estimates of the underlying 
CMB power  spectrum using the mode-mode coupling matrix corresponding to the  P06 mask 
following the MASTER approach~(\cite{Hivon2002,Hinshaw03,Tristram2005}). We  then remove beam and pixel effects from each of these $24$
full sky power spectra. Our final $EE$ power spectrum is simply an uniform average of these
$24$ cross-spectra. We rely upon Monte Carlo simulations to compute the error bars as well 
as possible bias in the extracted power spectrum. 

The neighboring multipoles in the power spectrum become coupled since
the spherical harmonics lose orthogonality on a masked sky. Hence, to
obtain full information about the two-point correlation function of
the resulting power spectrum one needs to construct the covariance
matrix,
$$
\left <\Delta C_l \Delta C_{l'}\right>= \left<( C_l-\left< C_l\right>)
(C_{l'}-\left<C_{l'}\right>)\right>\ ,$$  
We compute the covariance matrix by Monte Carlo simulations.
The correlations can be minimized by suitably binning the power spectrum.
We use a binning identical to that used by the WMAP team. Let $C_{\alpha}$ 
denote the binned power spectrum. Then the covariance matrix of the
binned spectrum is  obtained as $$\left <\Delta C_{\alpha} \Delta C_{\alpha'} \right>= \left<(C_{\alpha}-\left< C_{\alpha}\right>)( C_{\alpha'}-\left< C_{\alpha'}\right>)\right>\ .$$ The standard
deviation obtained from the diagonal elements of the binned covariance matrix 
gives the error-bars on the binned final spectrum. 
Since the cosmic variance of the CMB power spectrum
decays as $\sim 1/(2l+1)$, the diagonal terms in the above correlation
matrix decay with increasing multipoles. For a  visual comparison of correlation between different bins we
define a correlation matrix, $C_{\alpha \alpha'}$, of the binned power 
spectrum, where,
\begin{equation}
C_{\alpha \alpha'}=\frac{\left <\Delta C_{\alpha} \Delta C_{\alpha'}\right>}{\sqrt{\left<(\Delta C_{\alpha})^2 \right> \left<(\Delta C_{\alpha'})^2\right>}}\,.
\label{eq:correl_mat}
\end{equation}
All the elements of this matrix are bound to lie between $[-1,1]$. 

\section{Results}
\label{Result}
\subsection{Temperature Power Spectrum} 
The temperature power spectrum for $5$-year WMAP data is obtained using the
same procedure as described in~{\cite{Saha2008}).
The entire sky is divided into $9$ regions depending on the level of foreground 
contamination. 
The whole cleaning is done with the iterative method, starting from the dirtiest region. 

\begin{figure}[h]
\includegraphics[scale=0.45,angle=-90]{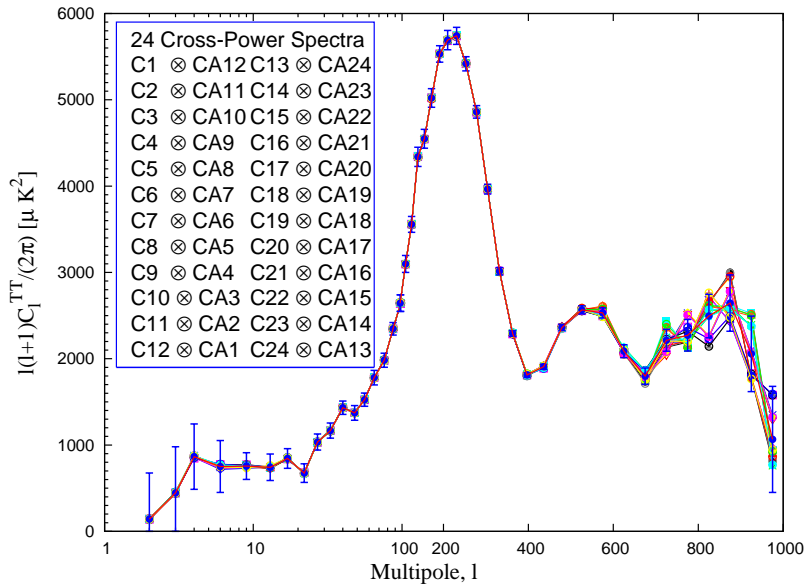}
\caption{
The 24, binned, $TT$ cross power spectra obtained by using the
WMAP 5 year data. All the different combinations of cleaned maps used are 
shown in the box. The average power spectrum along with error bars 
(blue points) is also shown. The red line joins the individual
binned averages. 
}
\label{fig:crosscl}
\end{figure}

All the 24 cross power spectra for temperature anisotropy are shown in 
 Fig. \ref{fig:crosscl}. We form an uniform average power spectrum  by averaging
over these 24 cross power spectra. While obtaining the cleaned power spectrum we use the 
KQ85 mask to remove the residual foreground contamination near the Galactic plane.

Even after applying the KQ85 mask which also removes a circular region around 
each of the known point sources, residual unresolved point sources cause 
a significant contamination in this power spectrum. There have been
several attempts to measure unresolved point source contamination in the 
CMB maps (~\cite{Teg98, Komatsu03, Huffenbe06}). We estimate unresolved 
point source contamination in our final power spectrum using the model
presented by~\cite{Nolta08} following an approach similar to that 
in~\cite{Saha2006,Saha2008}.

We compute the bias in the extracted spectrum by performing $150$
Monte Carlo simulations.  First we generate synchrotron, free-free and
thermal dust maps corresponding to different WMAP frequencies using
the Planck Sky Model\footnote{ A development version of the PSM can be
obtained upon request from the Planck Working Group 2, see
http://www.apc.univ-paris7.fr/APC\_CS/Recherche/Adamis/PSM/psky-en.php.}
(PSM).  
Although several options are available in the PSM to generate  galactic emission
(e.g. with or without spinning dust, with or without  small scales added), the largest scales in temperature are strongly  constrained by observation,
and the impact of the choice of a  particular model is not a major source of uncertainty. In our  simulations, we use a single set of galactic emission maps,
which  comprise a two-component dust model based on SFD model 7, synchrotron  map with varying spectral index in agreement with the first year
WMAP  data, and free-free emission with fixed spectral index, obtained from  an H-alpha template corrected for galactic dust extinction.
The exact polarisation properties of the galactic foregrounds, in  particular that of dust emission, are poorly constrained by observations.
For the present work, we use version 1.6.4 of the PSM  
(see Betoule et al. (2009) for details about the polarised galactic  emission).
 In the next step we
randomly generate CMB maps assuming the standard $LCDM$
model~(\cite{Spergel03}).  Each random realization of the CMB map is
then added to the combined mixture of all three foreground components
corresponding to the $5$ WMAP frequencies. Using the $5$ year beam
transform functions for different DAs provided by the WMAP science
team, we transform the $5$ resulting maps into $10$ maps.  Each map at
this step has a resolution appropriate for the corresponding DA.  We
then generate random noise maps corresponding to each detector. The
random noise maps are generated by sampling a Gaussian distribution
with unit variance and then multiplying each Gaussian variable by
$\sigma_{0}/\sqrt{N_{p}}$, where $\sigma_0$ is the noise per
observation~(\cite{Hinshaw08}) and $N_p$ the effective number of
observations at each pixel. The values of $\sigma_0$ depend on the DA,
with the smallest value for the K band DA and largest for the W band
DAs. Finally the noise maps are added to the CMB plus foreground maps
for different DAs. These maps with CMB signal, detector noise and
foreground are then passed through the same power spectrum estimation
method as in the case of observed data. The mean of the $150$
extracted spectra gives the final simulated power spectrum. The
standard deviation of the $150$ simulations gives the error. The
difference between the simulated power and the input $LCDM$ power
gives a measure of the bias in our method. This bias is subtracted
from the extracted power spectrum in order to get the final result.

The precise magnitude of the bias depends on the theoretical model with
which we compare our extracted power spectrum. 
In other words, before we compare our extracted power to a theoretical 
model, we must correct for bias using the corresponding model power spectrum.
Here we use the WMAP best fit $LCDM$ model to compute the theoretical power
spectrum.

The final temperature power spectrum using the Internal Power Spectrum
Estimation (IPSE) method, after correcting for bias, is shown in
Fig. \ref{fig:TT2}.  We find that it is in good agreement with the
WMAP best fit $LCDM$ model. The simulation results are also shown in 
Fig. \ref{fig:TT2}.
After bias correction we find the
quadrupole power ($l(l+1)C_l/2\pi$) equal to $532$ $\mu K^2$ compared
to the value of $213.4$ $\mu K^2$ estimated by the WMAP science team.
The quadrupole extracted from IPSE is, therefore, in much better
agreement with the theoretical model.


The correlation matrix, Eq. \ref{eq:correl_mat}, for temperature power spectrum 
is shown in Fig. \ref{fig:TTcovar}. We find, as expected, that the 
off diagonal matrix elements are negligible compared to the diagonal
elements.

\begin{figure}[h]
\includegraphics[scale=0.55,angle=-90]{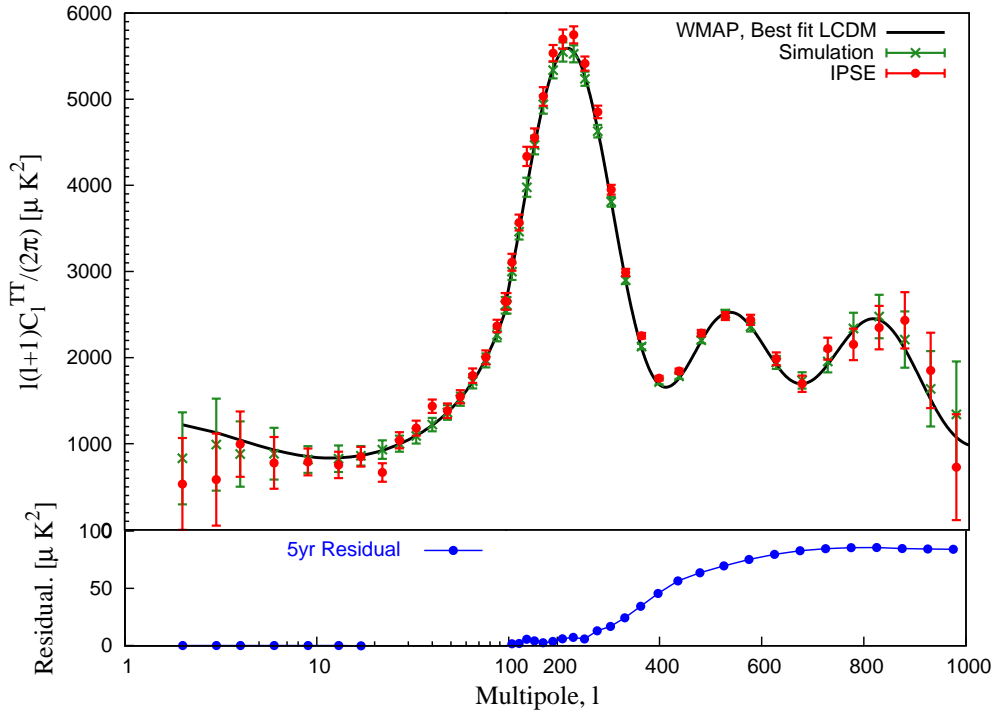}
\caption{The final, binned, 
$TT$ power spectrum using IPSE (red dots)  along with error bars
for the 5 year WMAP data,
after subtracting the bias extracted
using simulations. The error bars are also obtained from simulations. The
WMAP $LCDM$ best fit model (black line) and the simulation results (green
crosses) are shown for comparison. The bottom panel shows the correction made for residual power from unresolved point source
contamination.}
\label{fig:TT2}
\end{figure}

\begin{figure}[h]
\includegraphics[scale=0.8,angle=0]{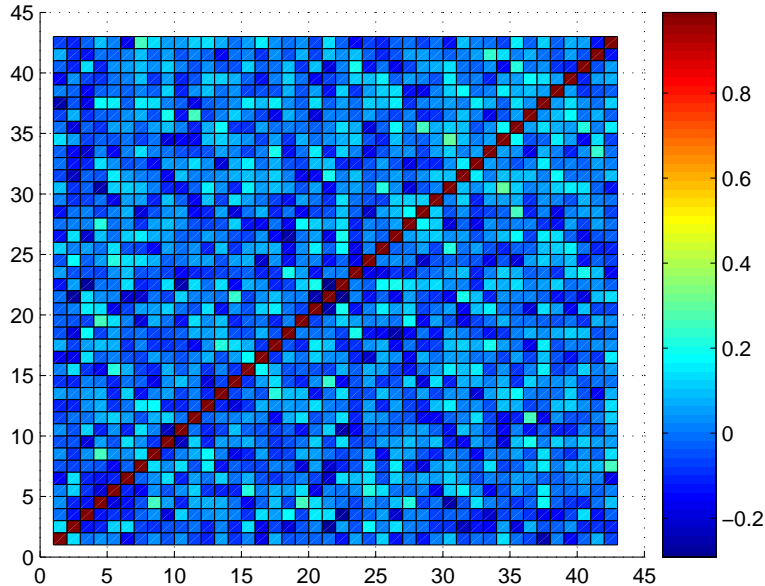}
\caption{ The correlation matrix elements, $C_{\alpha\alpha'}$, defined
in Eq. \ref{eq:correl_mat}, for the
temperature power spectrum, plotted with respect to the bin indices $\alpha, \alpha'$.}
\label{fig:TTcovar}
\end{figure}

\subsection{The $TE$ power spectrum}

The WMAP polarization CMB maps are cleaned using a single iteration rather
than the nine iteration procedure followed for temperature anisotropy. We 
apply the P06 mask right in the beginning. The mask removes
27\% of the entire sky region near the galactic plane. The cleaning algorithm
is applied only to regions outside the P06 mask. Hence we make no attempt to 
produce a full sky cleaned polarization map.

The error bars plotted in the $TE$  power spectrum are obtained
by Monte Carlo simulations. We generate $150$ random samples of data using the 
$LCDM$ model~(\cite{Spergel03}) and simulated foregrounds and detector noise. 
First we generate the synchrotron and thermal dust polarized foreground maps corresponding to 
different frequencies in terms of Q and U maps using the  
PSM version 1.6.4. The free-free emission 
is not polarized and hence not 
included.
The anomalous dust emission is also assumed to be unpolarized and thus
is not included either.
Using the
HEALPix \footnote{http://healpix.jpl.nasa.gov} command
{\tt synfast}
 we generate random realization of CMB polarization maps in terms of 
Q and U maps. 
The random CMB realization and foreground maps are smoothed
by the beam functions corresponding  to  ten different DAs. We next obtain E-mode polarization maps from these Q and U maps.
Then we generate random noise maps for each DA in terms of Stokes parameters Q and U using Cholesky decomposition technique for generating correlated Gaussian random variables using the WMAP supplied 
$2 \times 2$ QU intra-pixel noise covariance matrices.
These Q and U noise maps are converted to E mode noise maps. The final E-maps
including detector noise, foregrounds and CMB signal are passed through 
the same cleaning pipeline as the observed polarization data. 
In order to minimize the correlation among neighboring $l$ modes, the final 
power spectrum is binned in the same way as the WMAP 5 year result. Here 
also the standard deviation obtained from the diagonal elements of the binned covariance matrix is used as the error bars on the binned final spectrum extracted from the WMAP data.

The extracted $TE$ power spectrum along with the WMAP results and the best
fit $LCDM$ model is shown in Fig. \ref{fig:TE1}. 
The binned $TE$ power spectrum, using the same binning scheme as used by
the WMAP team, is shown in 
Fig. \ref{fig:TE2}. The error bars are computed by simulations.
The simulation results are shown in Fig. \ref{fig:TE2_sim}. We find
that the 
bias is small for all the bins. Only at small $l$, $l<10$, do we find a 
noticeable negative bias. For larger $l$, the bias is practically negligible.
The bias corrected $TE$ power spectrum 
is shown in Fig. \ref{fig:TE3}. The spectrum obtained by the WMAP science 
team as well as their best fit $LCDM$ model is also shown. We find
good agreement with the WMAP result. However we obtain slightly smaller 
error bars. We discuss the possible cause for this in the next subsection.
The correlation matrix elements, Eq. \ref{eq:correl_mat}, are shown in Fig. 
\ref{fig:TEcovar}. Here also we see that they are dominated by diagonal 
elements. 

\begin{figure}[h]
\includegraphics[scale=0.55,angle=-90]{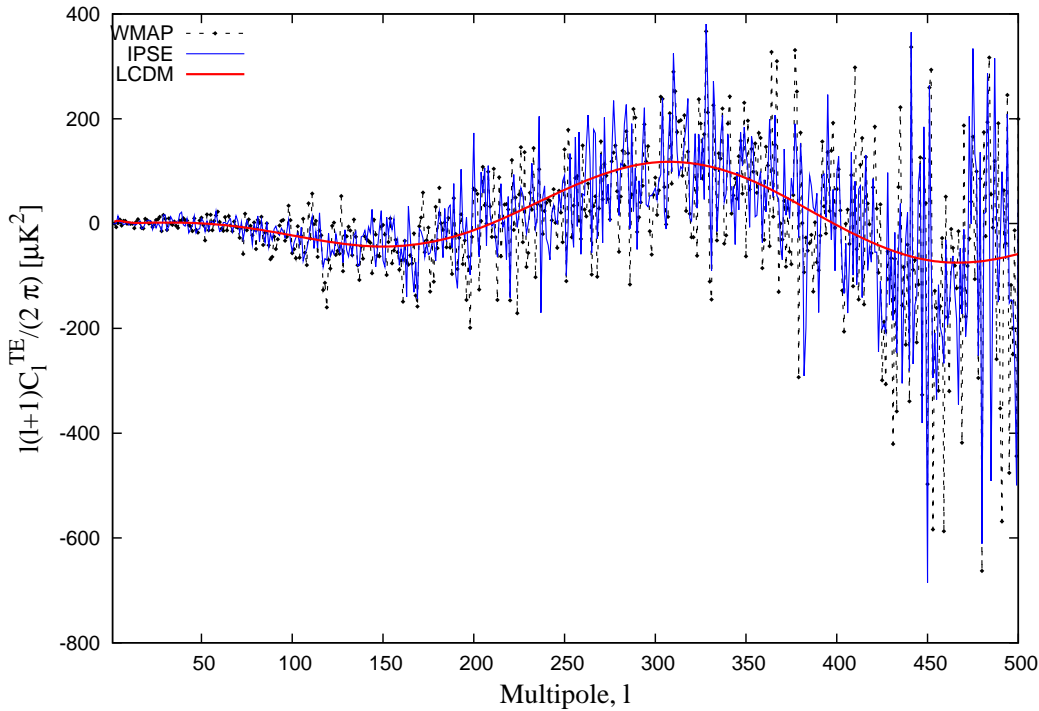}
\caption{The cleaned $TE$ power spectrum using IPSE (solid blue line)  
along with 
WMAP result (dashed black line). 
The WMAP best fit $LCDM$ power spectrum (thick solid red line) 
is shown for comparison. }
\label{fig:TE1}
\end{figure}

\begin{figure}[h]
\includegraphics[scale=0.55,angle=-90]{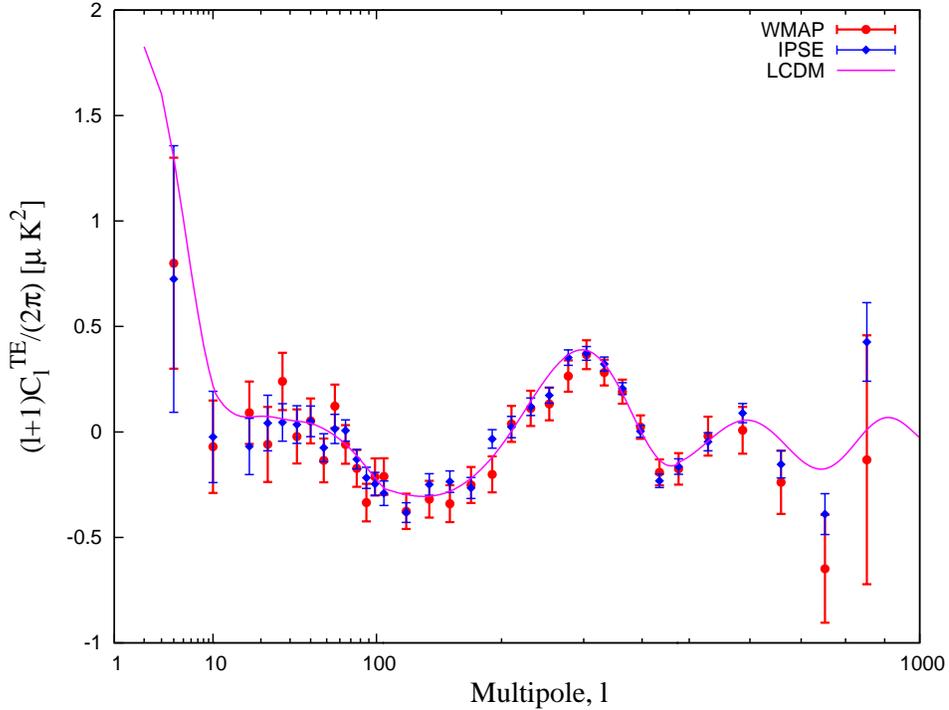}
\caption{The final binned $TE$ power spectrum using IPSE (blue diamonds) along 
with error bars, 
compared with the WMAP results (red dots). 
The WMAP best fit $LCDM$ result (solid pink 
line) is also shown. }
\label{fig:TE2}
\end{figure}

\begin{figure}[h]
\includegraphics[scale=0.55,angle=-90]{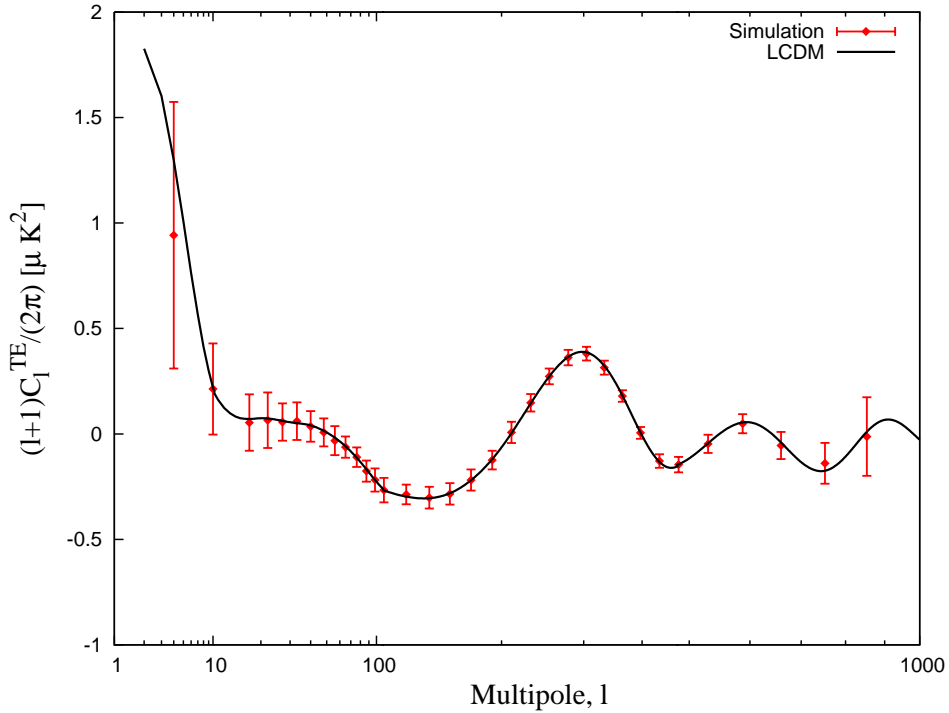}
\caption{The binned $TE$ power spectrum, along 
 with error bars, obtained from simulations (red dots).  
The input $LCDM$ model (solid black 
line) is also shown. }
\label{fig:TE2_sim}
\end{figure}

\begin{figure}[h]
\includegraphics[scale=0.55,angle=-90]{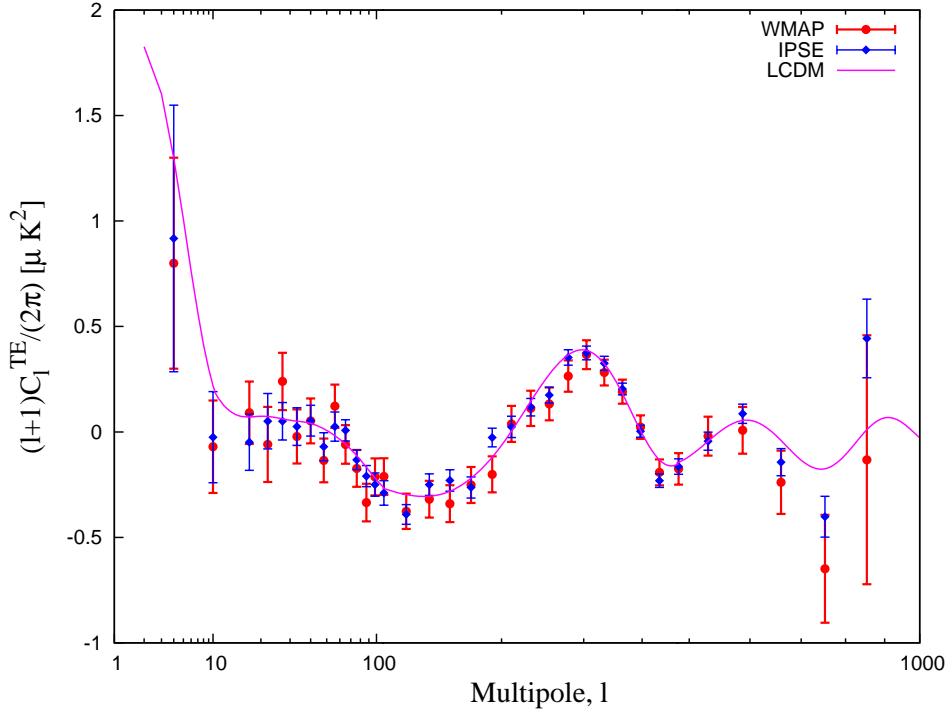}
\caption{The bias corrected, binned 
$TE$ power spectrum using IPSE (blue diamonds) along 
with error bars, compared with the WMAP results  (red dots).
 The WMAP best fit $LCDM$ power spectrum (solid pink line) is also shown.
  }
\label{fig:TE3}
\end{figure}

\begin{figure}[h]
\includegraphics[scale=0.80,angle=0]{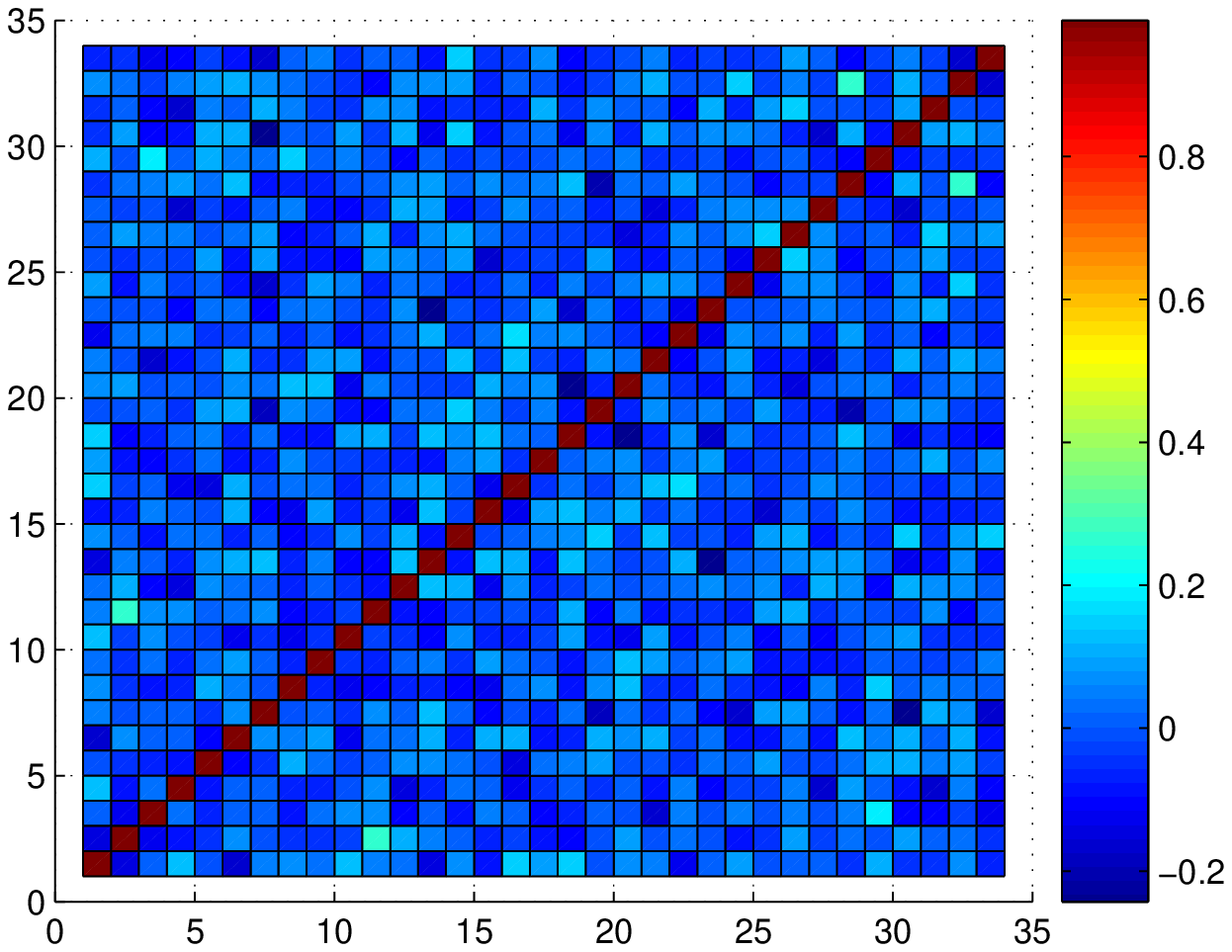}
\caption{The covariance matrix elements, $C_{\alpha\alpha'}$, defined
in Eq. \ref{eq:correl_mat}, for the
$TE$ power spectrum plotted with respect to the bin indices $\alpha,\alpha'$.}
\label{fig:TEcovar}
\end{figure}

\subsection{The $EE$ power spectrum} 
The binned $EE$ power spectrum, using the WMAP 
binning procedure, along with the simulation results, 
is shown in Fig. \ref{ee_ps_w_err} for the 5 year data. 
In this figure we also show the $EE$ power spectrum extracted by 
the WMAP team along with the WMAP best fit $LCDM$ model. 
In Fig. \ref{ee_ps_w_err} 
we follow exactly the WMAP binning procedure and consider
only the multipoles $l\ge 50$. The results for low multipoles $l<50$
are shown separately.
We find that our extracted power spectrum is in good agreement with 
that obtained by the WMAP team but with significantly smaller error bars.
Furthermore 
the binned simulated power spectrum 
is found to be close to the input $LCDM$ power over the entire multipole range.
Only at small $l$ do we find a significant positive bias. 
For the remaining multipoles
the simulation results match the input power within error bars.
The bias corrected power spectrum is shown in Fig. \ref{ee_ps_w_err1}.
We see that the bias corrected
spectrum is in reasonable agreement with the best fit $LCDM$ 
model.
The correlation matrix elements are shown in Fig. \ref{fig:EEcovar}.
We again find that the correlation matrix is dominated by diagonal matrix
elements. 

Our extracted power spectrum along with the simulation
results for low $l$ ($l\le 50$) are shown in Fig. \ref{ee_ps_lowl}. 
The WMAP power spectrum as well as their best fit $LCDM$ model is 
also shown. Here we have chosen the binning that was used by the WMAP team
for the TE power spectrum.
The bias corrected power spectrum is shown in 
Fig. \ref{ee_ps_lowl_bias_corrected_w_err}.

Since the power spectrum estimation methodology described in this work
is fundamentally different from WMAP team's approach we should not
expect both methods to produce identical error-bars on the derived
power spectra, although the power spectra themselves, obtained by
using the two methods, are in reasonable agreement with each other.
We find that our error-bars on the polarization power
spectra are smaller compared to those obtained by the WMAP science
team. This effect could be explained by noting that we use more
detector maps for polarization power spectrum estimation than WMAP
science team.  Using more detectors increases the signal to noise
ratio of the cleaned map by decreasing the effective noise level. The
reduced noise level leads to lower error-bar on our polarization power
spectra.  Furthermore, in the case of noisy
polarization data the weights tend to be inversely proportional to the
noise. Due to this inverse noise weighting each cleaned map has lesser
noise in comparison to the least noisy K or Ka band maps. This may be
another reason for smaller error bars in the case of polarization
power spectrum.

\begin{figure}[h]
\includegraphics[scale=0.55,angle=-90]{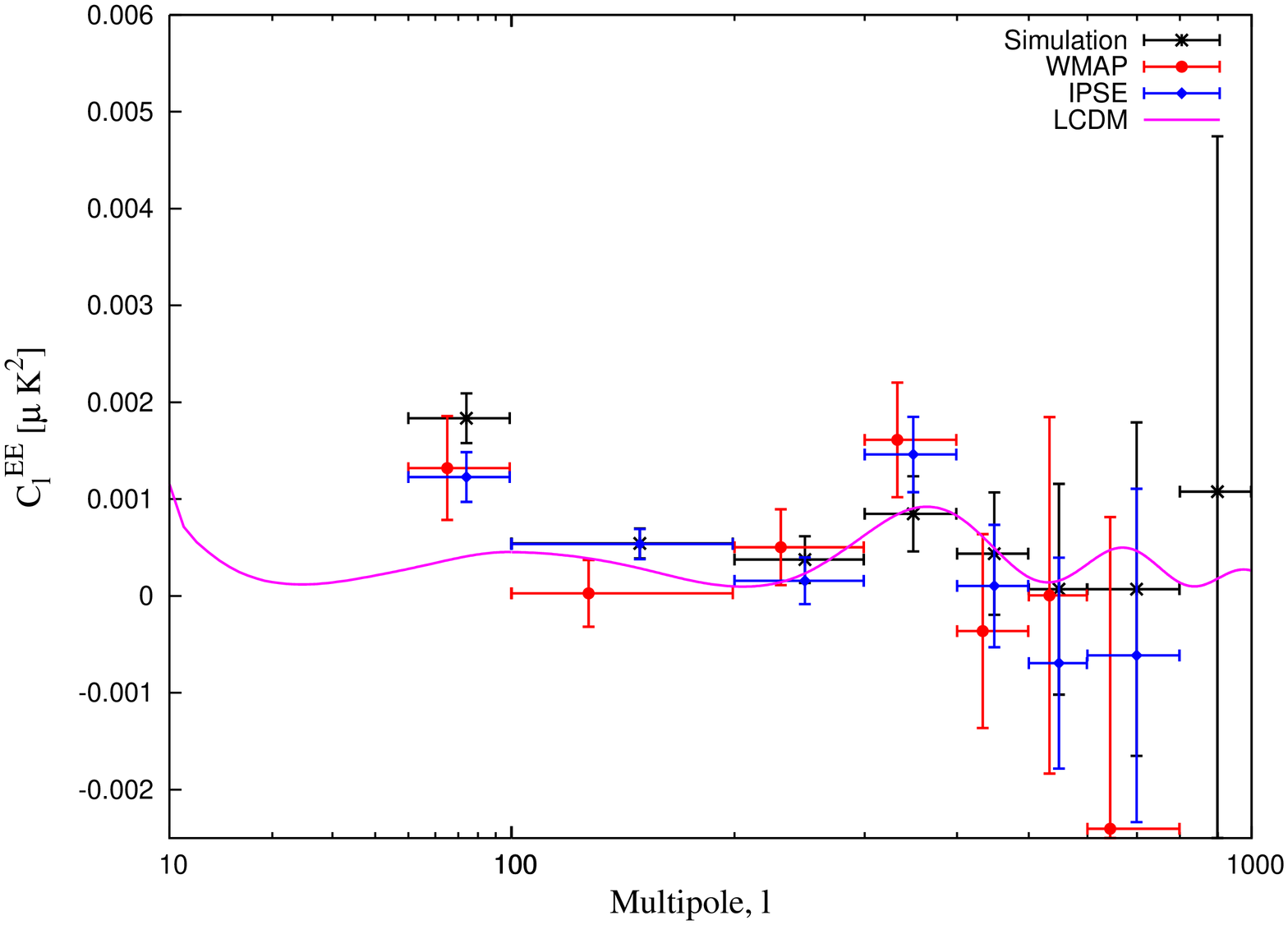}
\caption{The CMB polarization $EE$ power spectrum using IPSE (blue diamonds) 
along 
with error bars, compared with the WMAP result (red dots). The WMAP best fit 
$LCDM$ power spectrum (pink solid line) and
the simulation results (black crosses) are also shown. 
}
\label{ee_ps_w_err}
\end{figure}

\begin{figure}[h]
\includegraphics[scale=0.55,angle=-90]{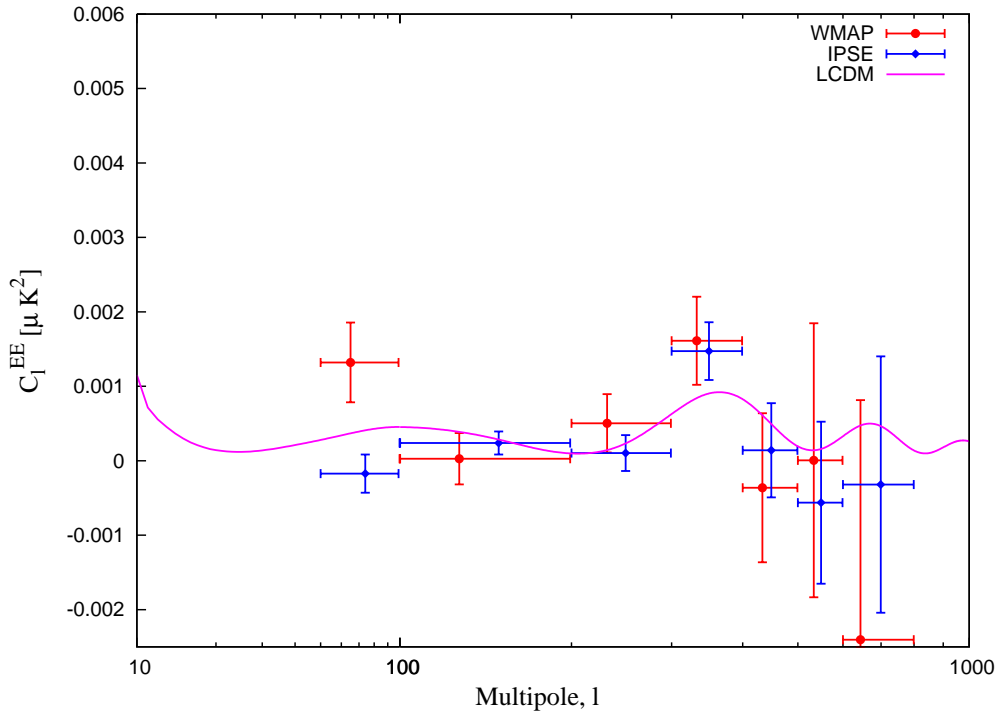}
\caption{The CMB polarization $EE$ power spectrum using IPSE (blue diamonds) 
along 
with error bars, after correcting for bias. The WMAP result (red dots) 
and the theoretical $LCDM$ spectrum (pink solid line) are 
shown for comparison. 
}
\label{ee_ps_w_err1}
\end{figure}

\begin{figure}[h]
\includegraphics[scale=0.80,angle=0]{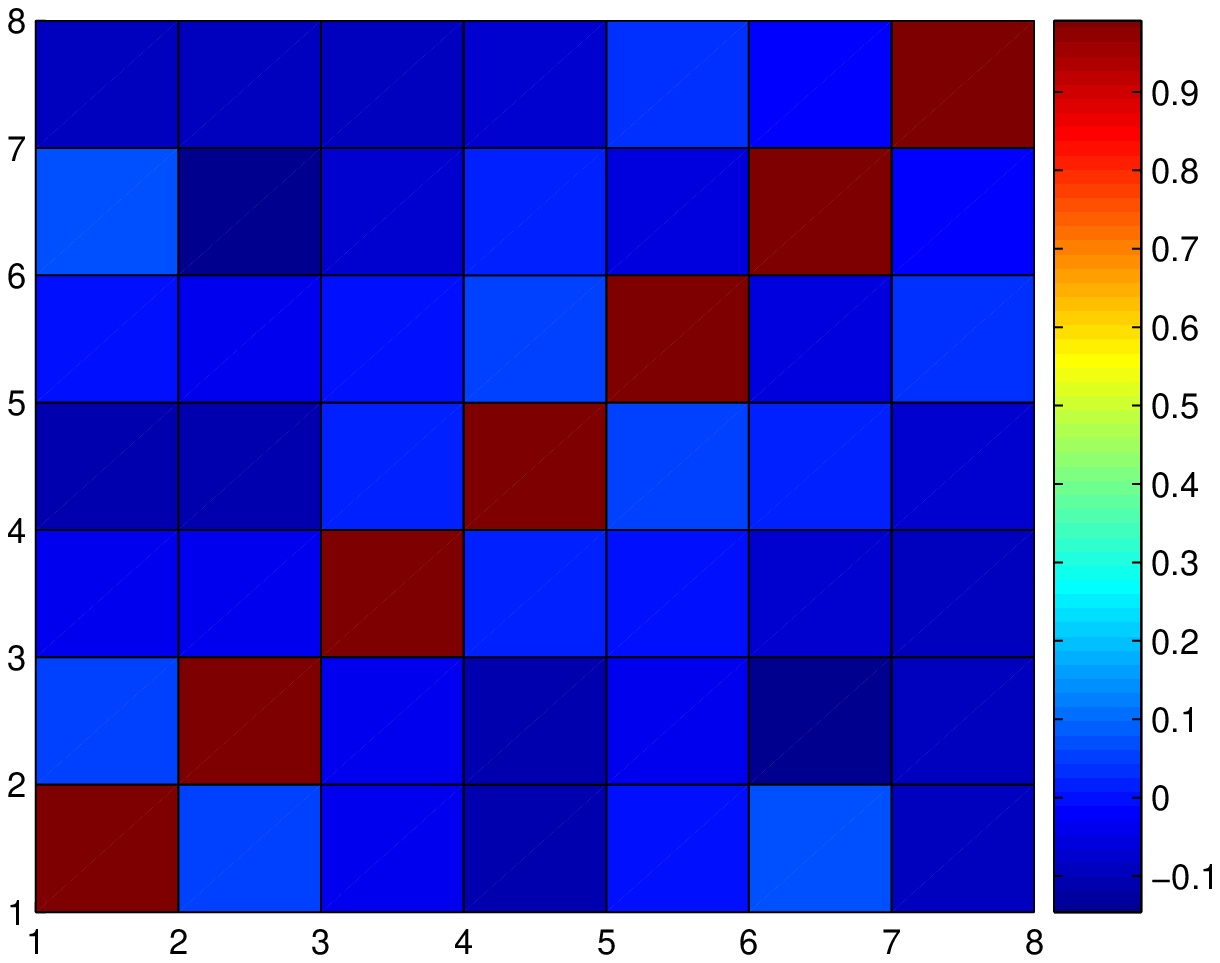}
\caption{The correlation matrix elements, $C_{\alpha,\alpha'}$, defined in
Eq. \ref{eq:correl_mat}, for the
$EE$ power spectrum plotted with respect to the bin indices $\alpha,\alpha'$.}
\label{fig:EEcovar}
\end{figure}

\begin{figure}[h]
\includegraphics[scale=0.55,angle=-90]{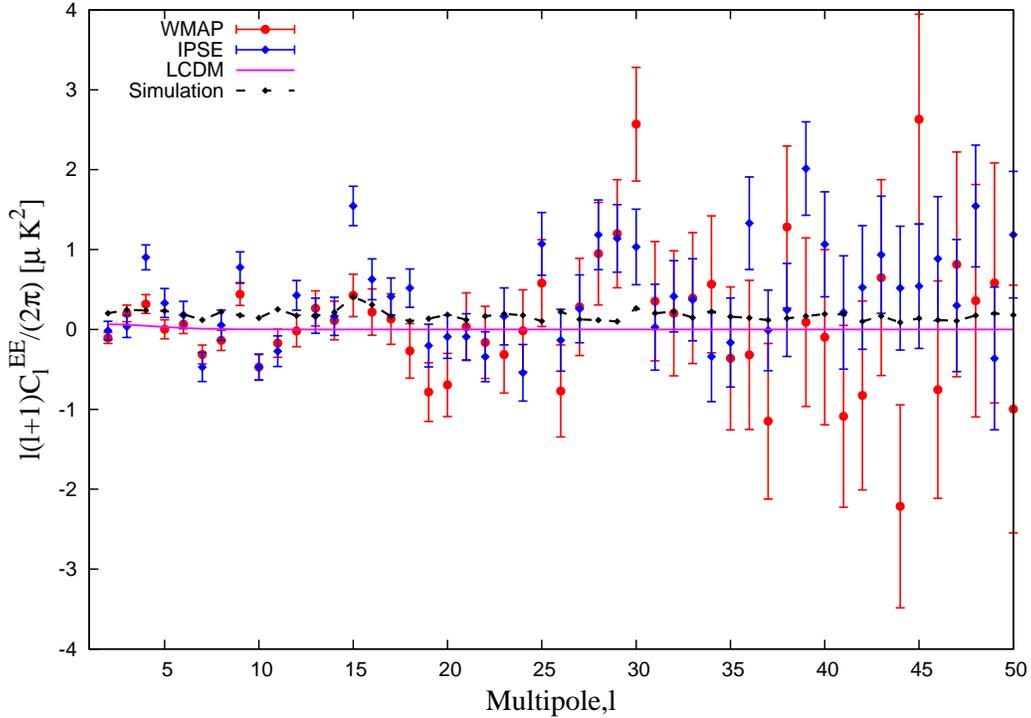}
\caption{The binned $EE$ power spectrum using IPSE (blue diamonds) 
at low-$l$ along with 
the results by the WMAP science team (red dots). The 
 theoretical $LCDM$ spectrum (solid pink line) and 
the ensemble averaged $EE$ power spectrum from simulated data (dashed
black line) are also 
shown.} 
\label{ee_ps_lowl}
\end{figure}

\begin{figure}[h]
\includegraphics[scale=0.55,angle=-90]{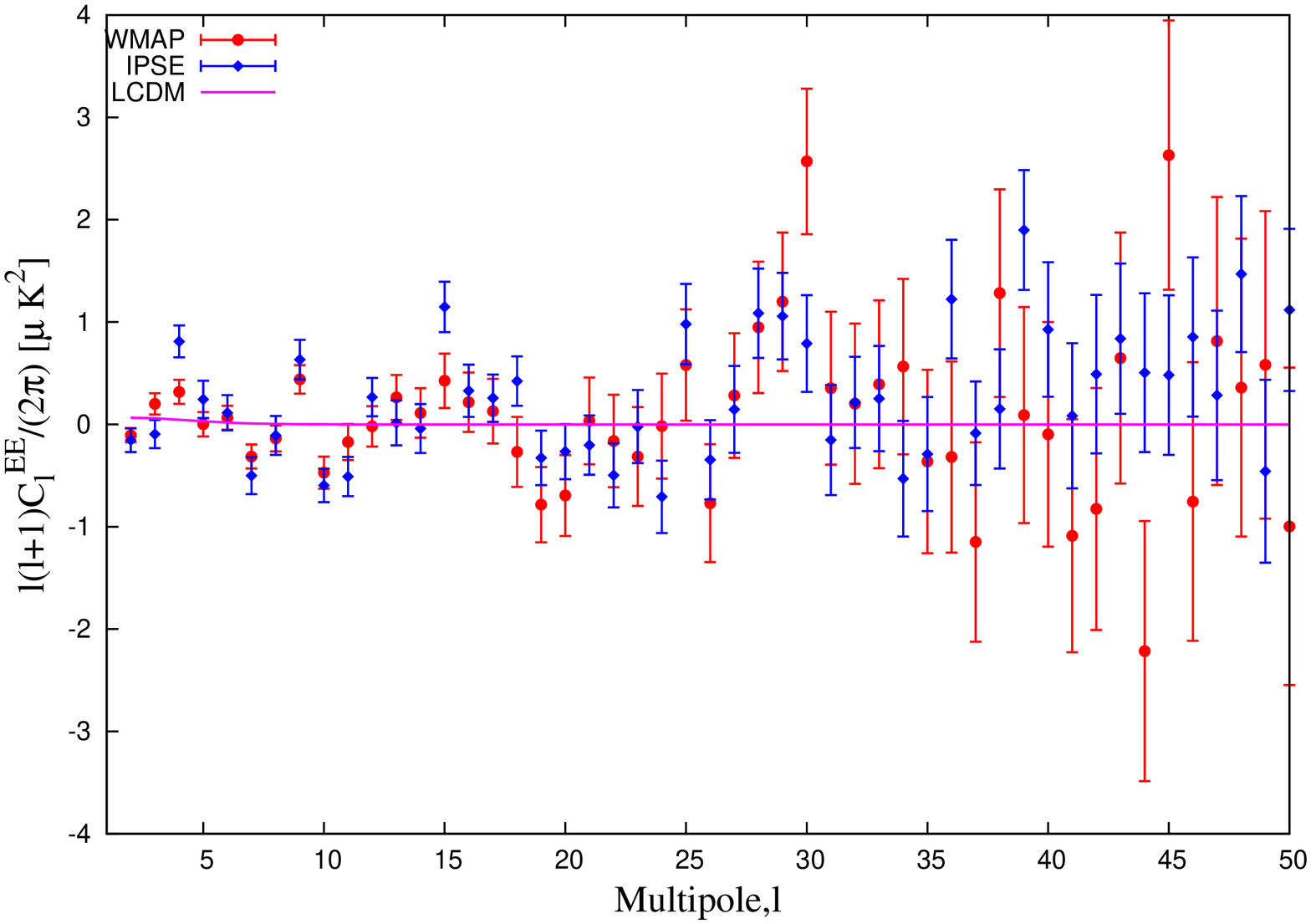}
\caption{The bias corrected $EE$ power spectrum using IPSE (blue diamonds) 
at low-$l$ compared with the WMAP results (red dots). 
The WMAP best fit $LCDM$ power spectrum (pink line) is also shown. } 
\label{ee_ps_lowl_bias_corrected_w_err}
\end{figure}

\section{Conclusion}
\label{Conclusion}
We have used a model independent method to estimate the CMB
temperature and polarization power spectrum using WMAP $5$-year data.
The method is based on the assumption that the CMB signal is
independent of frequency in thermodynamic temperature unit. Since the
foregrounds are frequency dependent in this unit, it is possible to
minimize the foreground power by making a linear combination of CMB maps
with a suitable choice of weights. For foreground minimization we use
the CMB maps in harmonic space. For the case of polarization, the raw
full-sky $Q$ and $U$ maps are first converted to $E$ and $B$ maps to
avoid any mixing of E and B modes. The total number of maps available
for each field for WMAP data is ten, corresponding to the ten DAs. We
create several cleaned maps by choosing different subsets of ten DAs,
such that each set contains only four DAs.  The detector noise power
is minimized by cross correlating cleaned maps obtained from distinct
DAs. This leads to a considerable reduction of the detector noise
power since, to a good approximation, noise is uncorrelated among
different detectors.

By utilizing all the ten WMAP DA maps to estimate the polarization
power spectra, we are able to provide more stringent constraints on
the spectra in comparison to that obtained by the WMAP science team. 
We find that the
error-bars on our $TE$ and $EE$ power spectra are smaller than those
obtained by the WMAP science team on all angular scales.  Another
possible reason for why we get smaller error bars in the case of
polarization power spectra is that in a noisy data the weights tend to
combine the maps in inverse noise weighted manner. This results in
each cleaned map having lesser noise than the least noisy K or Ka band
maps.

   In the case of $TT$ power spectrum
we find that our procedure does
   not remove all the unresolved point source contamination. This 
 contamination is significant at small angular scales where the detector 
 noise is also very large. Hence here our internal cleaning is not very 
 efficient. 
    The residual unresolved point source contamination
   is removed by using the WMAP point source model, as described
   in detail in \cite{Saha2006,Saha2008}. 

   We have performed detailed simulations of the $TT$, $TE$, $EE$
   power spectrum, using the WMAP best fit $LCDM$ model, along with 
   foreground and detector noise models, in order to determine if there
   exists any bias in the extracted power. In all cases 
   the bias is found to be  
   small for the entire multipole
   range. The extracted power, with or without bias correction, is found
   to be in good agreement with the WMAP results.
   In \cite{Saha2008} the authors 
   noticed a negative bias at low $l$ in the $TT$ power spectrum. 
   The negative bias arises due to
   a chance correlation between the CMB and the foregrounds. After correcting
   for the negative bias, we find that the quadrupole for the 
   WMAP 5 year data shows much better agreement with the $LCDM$ model, 
   in comparison to the result obtained by the WMAP science team. 
   Excluding $l=2$, we find negligible bias at all multipoles except  
   at very large $l$ values, where we find a small positive bias. 
   For the case of $TE$ power spectrum we also find a small negative bias at
   low $l$, $l<10$. For larger $l$ values the bias is negligible.
   For $EE$ power spectrum also the bias is found to be small compared to  
   the corresponding error bars. A significant
   positive bias is found only at low $l$.

To summarize, we have performed a completely independent reanalysis of
WMAP $5$ years temperature and polarization data. Our procedure uses
primarily the CMB data. Hence it is free from any bias that might
result from the
inadequacies and inaccuracies of the foreground modeling.  The
foreground templates and detector noise modeling is utilized only for
the purpose of bias analysis and error estimation. The bias is found
to be small for all the spectra over the entire multipole range. Our
results verify the basic power spectra results obtained by the WMAP
Science team.  We expect that the method will be very useful for
analyzing data from future CMB probes such as Planck.

\section{Acknowledgments}
We acknowledge the use of Legacy Archive for Microwave Background Data
Analysis. Some of the results of this work are derived using the publicly available
HEALPIx package (\cite{Gorski05}). (The HEALPIX distribution is publicly available from the 
website http://healpix.jpl.nasa.gov.) 
We acknowledge the use of the Planck Sky Model,
developed by the Component Separation Working Group (WG2) of the Planck
Collaboration. 
Pramoda K. Samal acknowledges
CSIR, India for financial support under the research grant CSIR-SRF-
9/92(340)/2004-EMR-I. A portion of the research described
in this paper was carried out at the Jet Propulsion Laboratory, California
Institute of Technology, under a contract with the National Aeronautics and
Space Administration. 
\bibliographystyle{hapj}
\bibliography{Polarization}

\end{document}